\documentclass[12pt]{iopart}
\usepackage{epsfig}
\usepackage{amssymb}

\newcommand{\be}{\begin{equation}}
\newcommand{\ee}{\end{equation}}
\newcommand{\bea}{\begin{eqnarray}}
\newcommand{\eea}{\end{eqnarray}}
\newcommand{\ba}{\begin{array}}
\newcommand{\ea}{\end{array}}

\begin{document}

\title[Michaelis-Menten dynamics out of a dichotomous ATP-based motor]{Michaelis-Menten dynamics of a polymer chain out of a dichotomous ATP-based motor}

\author{Alessandro Fiasconaro$^1$$^2$, Juan Jos\'e Mazo$^1$$^2$, Fernando Falo$^1$$^3$}

\address{$^1$ Departamento de F\'isica de la Materia Condensada, Universidad de Zaragoza, C/ Pedro Cerbuna, 12 50009 Zaragoza, Spain}
\address{$^2$ Instituto de Ciencia de Materiales de Arag\'on, CSIC-Universidad de Zaragoza, C/ Pedro Cerbuna, 12 50009 Zaragoza, Spain}
\address{$^3$ Instituto de Biocomputaci\'on y F\'isica de Sistemas Complejos, Universidad de Zaragoza, C/ Pedro Cerbuna, 12 50009 Zaragoza, Spain}

\ead{afiascon@unizar.es}


\begin{abstract} We present a model of an ATP-fueled molecular machine which push a polymer through a pore channel. The machine acts between two levels (working-waiting), and the working one remains active for a fixed time giving a constant force. The machine activation rate can  be put in relationship with the available ATP concentration in the  solution, which gives the necessary energy supply. The translocation  time shows a monotonic behaviour as a function of the activation  frequency and the velocity follows a Michaelis-Menten law that arises naturally in this description. The estimation of the stall force of the motor follows a corrected Michaelis-Menten law which still is to be checked in experimental investigation. The results presented agree with recent biological experimental findings.
\end{abstract}

\date{\today}


\section{Introduction}
In the last years there has been an important progress in the
experimental and theoretical study of transport mechanisms of
molecules inside cells and/or through cell membranes~\cite{RMP}. This effort is attracting nowadays more and
more attention from researchers in different scientific
disciplines. The increasing interest in this subject is related to the
intrinsic importance of understanding the basic mechanisms of the
living systems, but also to the enormous improvement of technological
capabilities at nanometer length scale. These, on the one hand, allow
to detect and to measure mechanisms at the nanoscale, and, on the
other hand, open the possibility of constructing from scratch
structures (with both natural and synthetic materials) able to imitate
the biological functioning \cite{mickler}. In this context, the
passage of biomolecules through nanopores is ubiquitous, both in
biological and nanotechnological processes. Examples of these two
types are the passage of mRNA through nuclear pores \cite{KasPNAS96}
or the translocation of DNA in graphene pores \cite{Gregory}.

In most cases, translocation is driven by constant fields in the pore
or by the difference in chemical potential between the two sides of
the membrane. However, in some cases the translocation is assisted by
an ATP-based molecular motor \cite{Mehta}. This is the case of DNA
bacteriophages in which the incoming DNA has to overcome a huge
pressure inside the virus capsid. This makes this kind of motors,
possibly, the most powerful of those known. In this type of motors ATP
hydrolysis is the fuel of the process.

In this paper we model the translocation process of a 1d chain pushed
by a molecular motor activated by ATP absorption. The motor is able to
drive with a constant force a polymer chain in one direction, while in
its activated state. The polymer diffuses freely otherwise
\cite{starikov,gomez}. The work reveals the Michaelis-Menten (MM)
behaviour of the polymer velocity and relates it with the MM enzymatic
reaction, according to a microscopic re-interpretation of the MM
kinetics which is a very actual topic of investigation \cite{walter,engl,mof}.

We will also study the behaviour of the motor against a pulling force,
the motor stall force, and its ability to package the polymer in a
finite region, as a first approach to a capsid effect.

The aim of this work is to present a simple model which captures the
main physical ingredients of the process. Remarkably, the model is able
to well describe some experimentally observed results and make new
predictions. The model could be also applied to different kinds of
systems. In this spirit, we do not pretend here to give a detailed
description of a particular molecular motor.

\section{The model}
\begin{figure}[b]
\centering
\includegraphics[width=10cm]{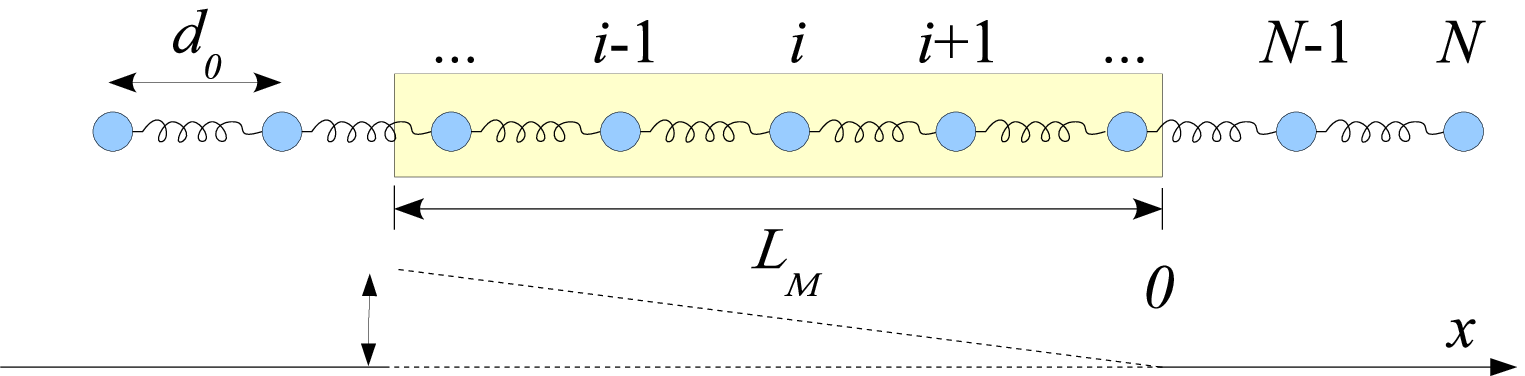}
\vskip -0.4cm
\includegraphics[angle=-90,width=10cm]{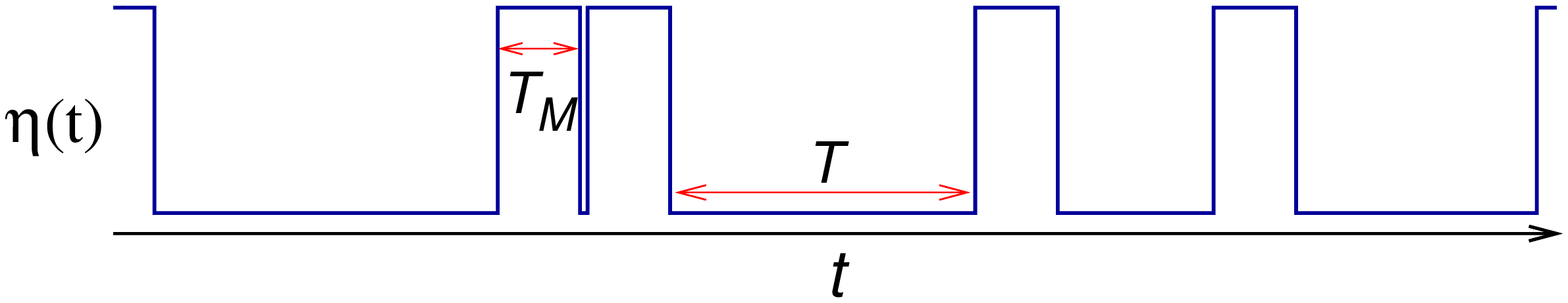}
\caption{Scheme of the dichotomous pushing force acting on a polymer
  chain compound by N monomers. $T_M$ is the working time of the
  motor, supposed fixed. $T$ is the mean waiting time in the inactive
  state. The motor acts on a region of length $L_M$, where
  the motor action can be expressed by the linear potential
  $-\eta(t)F_Mx$ (with $\eta=1$ for the active state and $\eta=0$ for the
  inactive one), indicated in the middle figure by two dotted
  lines. }
\label{schema}
\end{figure}

The polymer is modeled as a 1-dimensional chain of $N$ dimensionless
monomers connected by harmonic springs \cite{Rouse}. The total
potential energy is $ V_{\rm har}=\frac{k}{2}\sum_{i=1}^{N-1}
(x_{i+1}-x_i-d_0)^2,
 \label{v-har}
 $
\noindent where $k$ is the elastic constant, $x_i$ the position of
the $i$-th particle, and $d_0$ the equilibrium distance between
adjacent monomers. Along the work $d_0=1$ and $k=1$.

The translocation is helped by the presence of an ATP activated
molecular motor. It has spatial width $L_M$ and we set at $x=0$ its
right edge (see figure~\ref{schema}). The motor is characterized by a
fixed working time $T_M$ which follows the ATP absorption, which, in
its turn, occurs after a mean waiting time $T$, depending on the very
ATP concentration. In this sense the ATP molecules act on the motor as
a shot noise contribution able to switch on its activity. The motor
exerts a dichotomous force $F_M \eta(t)$ on the particles inside the
motor $(x \in [-L_M,0])$, where $\eta(t)$ is 1 during the working time
and 0 otherwise. As shown in \cite{Bust09}, the ATP
absorption follows an exponential distribution of waiting times: the
probability for an ATP adsorption in a time between t' and t'+dt'
after the last activity is proportional to $e^{-t'/T}$. Thus, the
activation probability of the motor is given by $P_{t'}= 1-e^{-t'/T}$
with $t'$ the motor residence time in its inactive state.

The dynamics of the $i^{\rm th}$ monomer of the chain is described by
the following overdamped Langevin equation: \be \dot{x}_i = -V_{har}'+
F_M\eta_i(t) + \xi_{i}(t),
\label{dyn}
 \ee where the damping (considered the same for all the monomers,
 which feel the viscosity independently on each other) is included in
 the normalized time. In the above dimensionless equation
   $d_0$ is taken as the unit of length, the force is expressed in
   units of $k d_0$, and the energy in units of $k d_o^2$.
 $\xi_{i}(t)$ are the Gaussian uncorrelated thermal fluctuations which
 follow the usual statistical properties $\langle\xi_i(t)\rangle=0$
 and $\langle\xi_i(t)\xi_j(t+\tau)\rangle = 2 D
 \delta_{ij}\delta(\tau)$ with $(i,j=1...N)$.

\section{Energy supply} The energy used by the motor to provide the
driving is given by the ATP hydrolysis.  In our model, each ATP
hydrolysis just activates the motor for a fixed working time $T_M$.
This description avoids more complex details which are
beyond the spirit of our work.

As a consequence of the ATP absorption, the motor changes its
conformational state and exerts a mean force $F_M$ during a fixed time
$T_M=1/\nu_0$. After that time the motor returns to its inactive
state. A new force is applied when the next ATP suitable quantity is
absorbed. It happens after a mean time $T=1/\nu$, which follows an
exponential distribution of waiting times. That way in our description
the activation frequency is proportional to the ATP concentration
\cite{sancho}: $\nu \propto [ATP]$. The hypothesis that the motor
works for a fixed time and that the statistics of the arrival of the
ATP molecules happens in an exponential distribution appears realistic
in good approximation as put in evidence by different experimental
works \cite{engl,mof,Bust09,Bust01,Bust05}.

With this definition the Michaelis-Menten law arises naturally from
the model. The motor duty ratio, the fraction of time that it is
active, is given by \be \frac{T_M}{T_M + T}=\frac{\nu}{\nu_0+\nu} =
\frac{[ATP]}{k_M+[ATP]}, \label{MM} \ee which is a Michaelis-Menten
law. The only hypothesis included in the derivation of the last
equality is that the motor activation rate $\nu$ is proportional to
the ATP concentration, being $k_M=[ATP]\nu_0/ \nu$ the Michaelis
constant.

The relationship between the mechanical description presented and the
Michaelis-Menten law is deeper than only the statistical ansatz $\nu
\propto [ATP]$. In the Michaelis-Menten enzymatic reaction, \be E + S
\longrightarrow \!\!\!\!\!\!\!\!\! ^{k_1} \ \ Z \longrightarrow
\!\!\!\!\!\!\!\!\!  ^{k_2} \ \ E + P, \ee the rate $k_1$ represents
the probability to form the compound $Z$ per unit of time and per unit
of $S$ ([ATP]), and $k_2$ gives the probability to form the product
$P$ per unit of time. In our mechanical and individual case (single
motor and single ATP event) $Z$ represents the ATP bound to the motor,
which occurs with the frequency $\nu$ ($\sim k_1 [ATP]$), while $P$
represents the motor action, which is completed within a time $T_M
=1/\nu_0$ ($\nu_0 \sim k_2$).  These relationships are in agreement
with the definition of the Michaelis constant $k_M=k_2/k_1$.

\section{Results}

We have numerically solved the Langevin equation~(\ref{dyn}). We are mainly
interested in the behaviour of the translocation time $\tau$ and
velocity $v$ as a function of the motor activation frequency $\nu$. We
will compare our results to some theoretical predictions, derived
below. For every computed point we performed at least $N_{exp}=10,000$
numerical experiments (for low frequency up to $50,000$) using a
stochastic Runge-Kutta algorithm with a time step $dt=0.01$. The
polymer is compound by $N$ monomers and starts with all the spring at
the rest length and the last monomer of the chain at $x_N=0$, just at
the exit of the motor. The noise intensity is held fixed at the
value $D=0.01$ and the intensity of the force is $F_M=0.1$. We have
chosen $L_M/d_0=5.5$.

{\em Polymer translocation.} We start studying the mean translocation
time and velocity of the polymer driven by the motor. With
respect to the polymer velocity, by summing up the $N$ terms of
equation~(\ref{dyn}) and averaging in time, we obtain for the centre of
mass of the chain the mean velocity $v$:
\be
\label{mmeq} v=\frac{F_M}{N}\sum_{i=1}^N \langle \eta_i(t) \rangle
= \frac{F_M}{N} \frac{n_{\rm mot}^{\rm on} (\nu)}{1+\nu_0/\nu}.
\ee
Here $n_{\rm mot}^{\rm on}(\nu)$ is the mean number of
monomers inside the motor during its activity, a number which is
expected to depend weakly on $\nu$.  Thus, the mean velocity of the
polymer depends on the force felt by the $n_{\rm mot}^{\rm on}$
monomers inside the machine which operates for fraction of time
$\nu/(\nu_0 + \nu)$, and the polymer shows a Michaelis-Menten law for
the velocity weakly moderated by the function $n_{\rm mot}^{\rm
 on}(\nu)$. Note that this velocity goes to zero as
$1/N$ for a large chain as expected for a motor which acts on a small
number of monomers and a polymer which moves in a dissipative media.

\begin{figure}[t]
\centering
\includegraphics[angle=-90, width=11cm]{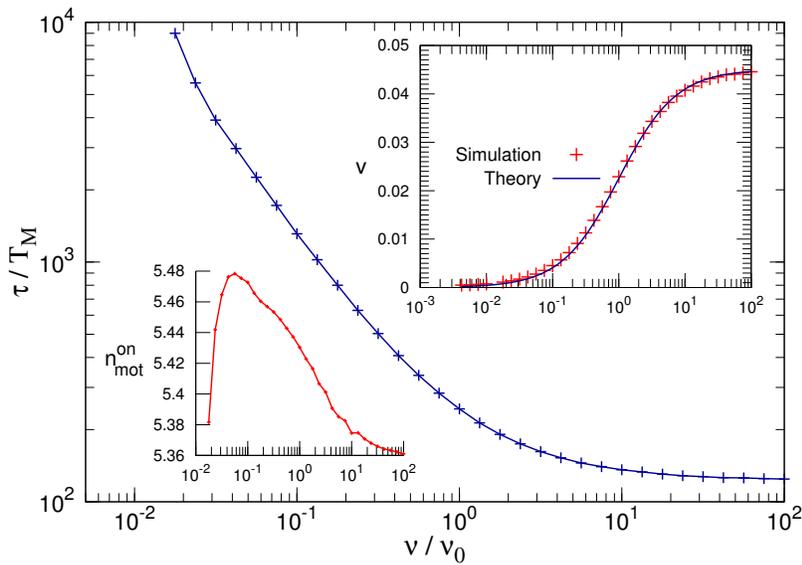}
\caption{Log-Log plot of the mean first passage time
  $\tau$ as a function of the mean frequency of the fluctuating
  force. Upper inset: mean velocity, following the Michaelis Menten
  law given by equation~(\ref{eqv}). Bottom inset: mean number of particle
  inside the motor during its active state. The thermal noise
  intensity is $D=0.01$ and $k=1$.}
\label{RA}
\end{figure}

We also want to mention that equation~(\ref{mmeq}) it is also valid in
other interesting situations too. For instance, systems with random
distributions of $T_M$ values with small dispersion around its mean
value or the case of a minimum threshold time for the motor
activation, which effectively imposes a maximum cutoff frequency in
the system (in this case $1/\nu=T_{thres}+T$).

Figure~\ref{RA} shows the main observables of the system and their
frequency dependence. The translocation time $\tau$ is computed as the
mean first passage time (MFPT), i.e. the average over the $N_{exp}$
realizations of the time spent by the centre of mass of the chain to
reach the position $x=0$. It is observed that this time decreases
monotonously as $\nu$ increases and reaches a limit value for large
enough values of $\nu$ ($T$ goes to zero and the motor is `on' most
of the time).  The relation between translocation velocity $v$ and
time $\tau$ is not trivial at all, as can be seen
in~\cite{fjf-sin} and~\cite{fjf-rtn}.

The physics of the problem is regulated by the mean number of monomers
in the motor during the working time, $n_{\rm mot}^{\rm on}$, a number
directly related with the elasticity of the chain. As seen in the
inset, for the used parameters this number is close to $L_M/d_0$ and
does not change importantly in all the frequency range covered. Thus,
as predicted by equation~(\ref{mmeq}) the polymer velocity follows a
Michaelis-Menten law (see inset in figure~\ref{RA}): \be v \simeq
v_{HR}^0 /(1+\nu_0/\nu) \label{eqv} \ee with $v_{HR}^0$ the high
frequency limit ($v_{HR}^0$=0.0450 in the plotted case, which is close
to 0.0458=$F_M n_M/N$ with $F_M=0.1, N=12$ and
$n_M=L_M/d_0=5.5$). Equation~({\ref{eqv}}) allows for a direct
experimental fit once the velocity at high ATP concentration (the high
rate limit of our system) is measured. With respect to the low
frequency limit, we can see from the equation that $v$ goes to
zero as $v \simeq v^0_{HR} (\nu/\nu_0) \simeq (F_M n_M/N)
(\nu/\nu_0)$.

\begin{figure}[tb]
\centering
\vskip -0.0truecm
\includegraphics[angle = -90, width=11cm]{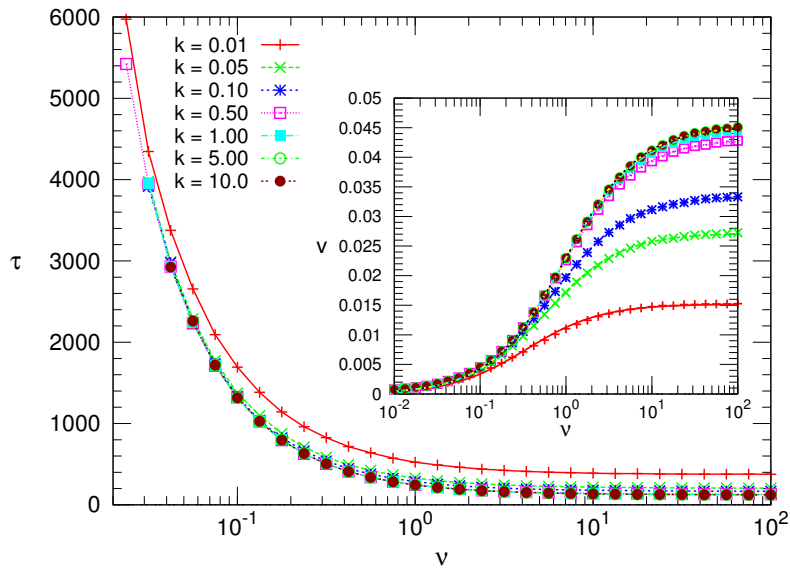}
\caption{Translocation time and velocity (inset) for different values
  of the elasticity of the chain $k$. Lines in the inset show the fit to
  $v=v_{HR}^0/(1+b\nu_0/\nu)$ where $b=1$ for $k>0.5$.}
 \label{ka}
\end{figure}

\emph{Elastic Constant.} In order to better understand how the
elasticity of the chain acts on the translocation dynamics, we have
computed the mean velocity and first passage time ($v$ and $\tau$) for
different values of the elastic constant between monomers $k$ (see
figure~\ref{ka}). We notice that similar behaviour is observed for the
different values considered. As expected, a smaller $k$ gives a lower
velocity, the polymer enlarges and the mean number of monomers in the
machine diminishes. At low $k$, $n_{\rm mot}^{on}$ changes
significantly with $\nu$ and $v$ shows deviations with respect to the
predictions of equation~(\ref{eqv}) but can be nicely fit by a
slightly modified law (see caption of figure~\ref{ka}). For high
$k$ ($k>1$) we reach the rigid chain limit of the system.

\begin{figure}[b]
\centering
\vskip -0.0truecm
\includegraphics[width=10cm]{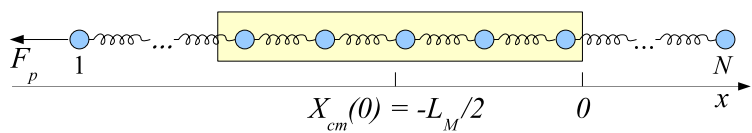}
\caption{A pull force $F_p$ is applied at the chain to measure the
  stall force of the motor.} \label{stall2}
\end{figure}

{\em Translocation in the presence of a pull force.}  A set of
simulations has been performed by applying a pull force $F_p$ on the
left extremum of the chain, acting against the motor (see
figure~\ref{stall2}). The initial condition is set with the polymer
centre of mass in the centre of the machine. The velocity of the
centre of mass is measured waiting for the exit on the left or on the
right of the potential region.

In this case, the mean velocity is given by
 \be \label{mmeq1} v=
   \frac{F_M}{N} \ \frac{n_{\rm mot}^{\rm on}
   (\nu;F_p)}{1+\nu_0/\nu}-\frac{F_p}{N},
 \ee
where the last term $F_p/N$ modifies equation~(\ref{mmeq}) by
taking into account the presence of an external force acting against
the machine. There we emphasize that the presence of a
pulling force also modifies the value of $n_{\rm mot}^{\rm on}$ with
respect to the unforced value.

Assuming again a weak dependence on $\nu$ and $F_p$ for $n_{\rm
  mot}^{\rm on}$ we can rewrite the previous equation as
\be \label{mmeq2} v \simeq
\frac{v_{HR}^p}{1+\nu_0/\nu}+\frac{v_{LR}^p}{1+\nu/\nu_0} \ee where
$v_{LR}^p=-F_p/N$ is the low frequency limit of $v$ and $v_{HR}^p$ the
high frequency one, $v_{HR}^p=(F_M n_{\rm mot;HR}^{\rm on}-F_p)/N$.

\begin{figure}[tbp]
\centering
\includegraphics[angle=-90, width=11cm]{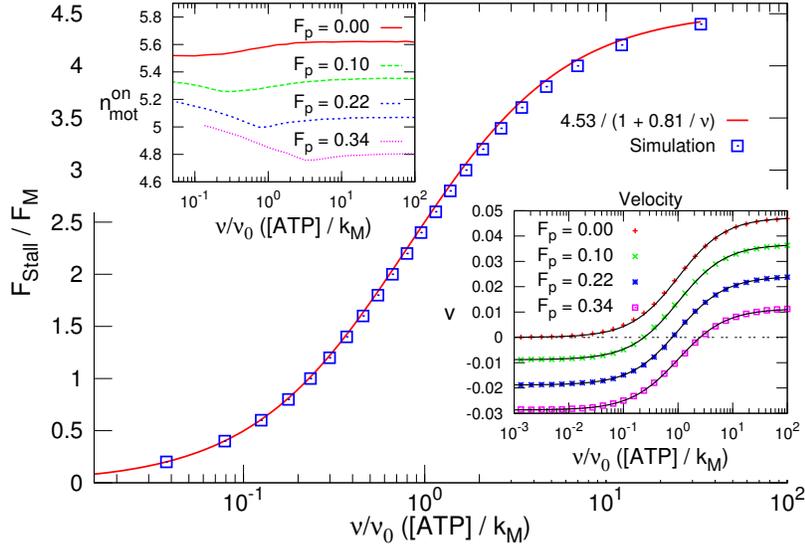}
\caption{Stall force as a function of the frequency of the dichotomous
  driving and prediction of equation~(\ref{eureka}). The insets show,
  for four values of $F_p$, $n_{\rm mot}^{\rm on}(\nu)$ (upper inset)
  and the mean velocity (lower inset) computed (symbols) and predicted
  (lines) by equation~(\ref{mmeq2}).}
\label{StallForce}
\end{figure}

Figure~\ref{StallForce} shows (lower inset) our numerical
results for the polymer velocity in the presence of a pull and it
allows for a comparison against the theoretical predictions.  As seen
in the figure, the polymer mean velocity can be understood in terms of
equation~(\ref{mmeq2}). Figure~\ref{StallForce} shows the excellent agreement
between this equation and the numerical calculations once the low and
high ATP concentrations (low and high frequency) values are
determined. The value of $v_{HR}^p$ present in that equation can be
evaluated experimentally by using the values of the velocity curves
for high ATP concentrations \cite{Bust05}. A rough estimation can be
also given by $v_{HR}^p \simeq (F_M L_M/d_0-F_p)/N$.

We have also studied the frequency dependence of $n_{\rm mot}^{\rm on}(\nu;F_p)$.
As visible in the top inset of figure~\ref{StallForce}, this quantity depends
weakly on $\nu$ for all the pulling forces and decreases slightly with
$F_p$ for all the frequencies. Then the velocity of the polymer shows
the expected Michaelis-Menten type behaviour for all the different pull
forces used in the calculations ($F_{p} = 0.1, 0.22, 0.34$).  We have
verified numerically that we can consider $n_{\rm mot}^{\rm
  on}(\nu,F_p) \simeq n_{\rm mot;HR}^{\rm on}(F_p) \simeq n_{\rm
  mot;HR}^{\rm on}(0)-aF_p$. In our case $a \simeq 2.35$ and $n_{\rm
  mot;HR}^{\rm on}(0) \simeq 5.59$, obtained through a fit procedure
of $n_{\rm mot;HR}^{\rm on}(F_p)$, the high rate values of $n_{\rm
  mot}^{\rm on}(\nu,F_p)$.

{\em Stall Force.} We will study now the stall force $F_{stall}$ of
the system, which is the value of the pull force $F_p$ for which the
polymer velocity is zero. In formulas

 \be \label{sf}
   F_{stall} = F_p (v=0) = F_M \frac{n_{\rm mot}^{\rm
       on}(\nu,F_p)}{1+\nu_0/\nu} \simeq F_M \frac{n_{\rm mot;HR}^{\rm
       on}(F_p)}{1+\nu_0/\nu}.
 \ee
Here we have used the weak frequency dependence of $n_{\rm mot}^{\rm
  on}$ as suggested by the inset of figure~\ref{StallForce}. A simple
approximation for $n_{\rm mot;HR}^{\rm on}(F_p)$ is to assume $n_{\rm
  mot;HR}^{\rm on}=L_M/d_0$. As discussed above, a more realistic one
is to take the linear dependence $n_{\rm mot;HR}^{\rm on}(F_p) \simeq
n_{\rm mot;HR}^{\rm on}(0)-aF_p$. Substituting the latter expression
in equation~(\ref{sf}), we obtain
 \be F_p = F_M \frac{n_{\rm mot;HR}^{\rm
    on}(0)}{1+\nu_0/\nu} - F_M \frac{aF_p}{1+\nu_0/\nu}.
 \label{eu1}
 \ee
By solving with respect to $F_p$, and introducing $F_{stall}$, we
obtain
 \be
     F_{stall} \simeq \frac{F_{stall}^{HR}}{1+b\nu_0/\nu}
 \label{eureka}
 \ee
where $b=1/(1+aF_M)$ and $F_{stall}^{HR} \simeq b F_M n_{\rm
  mot;HR}^{\rm on}(0) $ is the high frequency stall force.

Figure~\ref{StallForce} shows our numerical results for the polymer
stall force problem and compares them to our theoretical
predictions. We find an excellent agreement with the predictions of
our equation~(\ref{eureka}), where $b \simeq 0.81$ and $F_{stall}^{HR}
\simeq 0.453$. These numbers results from the values of $a \simeq
2.35$ and $n_{\rm mot;HR}^{\rm on}(0) \simeq 5.59$ obtained above..

In this model, the stall force is \textit{not weakly} dependent on the
frequency $\nu$, but changes from $0$ to $4.5 F_M$ in the range of
rate variation of figure~\ref{StallForce}. Thus an experimental
investigation can easily verify the stall force frequency dependence
by lowering the ATP concentration in the surroundings of the
motor. This way the working model can be verified and tested, and also
compared with different models. \footnote{We refer in particular to
  the outcomes of a similar motor model assisted by both a sinusoidal
  time dependence \cite{fjf-sin}, and a pure dichotomous driving
  \cite{fjf-rtn}.  Besides the not trivial behaviours of $F_{Stall}$ as
  a function of the frequency, a very weak variation of its values is
  there observed.}

\begin{figure}[]
\centering
\includegraphics[angle=-90,width=11cm]{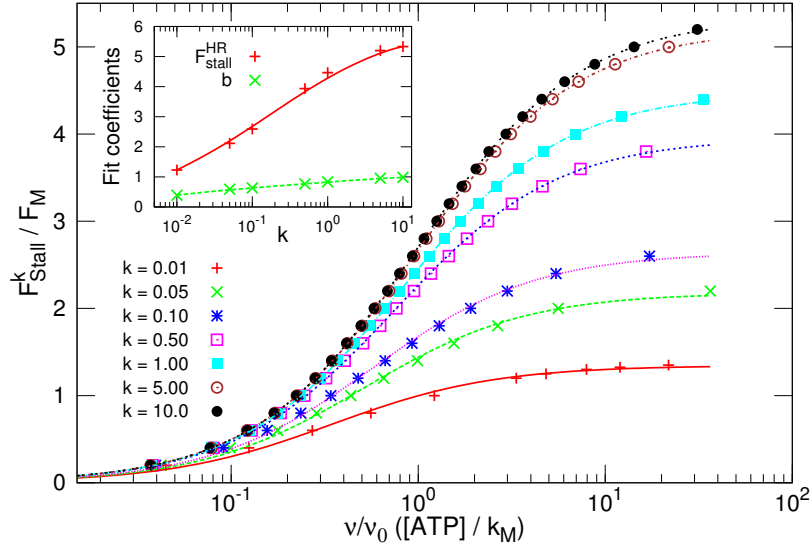}
\caption{Stall force $F_{stall}^k$ for various values of the elastic
  parameter $k$. Lines are the best fit according to equation
  (\ref{eureka}) and the inset shows the values of the fit parameters
  for the different $k$.}
\label{stall-k}
\end{figure}

{\em Stall force for different $k$.}  We have made a number of
simulations to study the dependence of the stall force with the
elasticity constant $k$. The results are plotted in
figure~\ref{stall-k} where the stall force is drawn as a function of
the frequency $\nu$ for various values of $k$. We can observe that the
stall force increases with $k$ and shows a saturation behaviour at high
$k$ (see the curves for $k=5$ and $k=10$) similar to the one of the
velocity (inset of figure~\ref{ka}).

The MM behaviour predicted by equation~(\ref{eureka}) is observed for
all the curves.  The inset of the figure shows the behaviour of the
coefficients $F_{stall}^{k, HR}$ and $b$ of the equation. As expected,
the rigid chain (high $k$) follows an exact MM law, $b=1$ in
equation~(\ref{eureka}).

{\em Polymer packing.} One of the recent most relevant
activity in the field is the study of the translocation
features in the DNA packing problem driven by molecular motors
\cite{Bust09,Bust01,Bust05}. One example is the $\phi 29$, a
bacteriophage virus which is able to inject his DNA in a bacteria in
order to replicate, and then repack it in his capsid. Remarkable
experiments \cite{Bust01} have measured the force of the motor as a
function of the number of monomers entered in the capsid. The model
here depicted is able to qualitatively reproduce the results there
reported.
\begin{figure}[bp]
\centering
\vskip -0.3 truecm
\includegraphics[width=10cm]{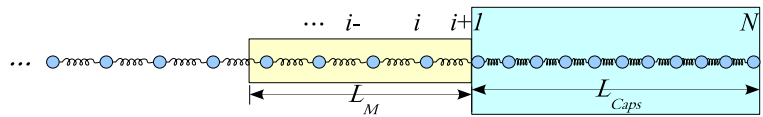}
\caption{Scheme of the motor pushing the polymer
inside the capsid of length $L_{Caps}$.} \label{caps}
\end{figure}
\begin{figure}[t]
\centering
\includegraphics[angle=-90, width=11cm]{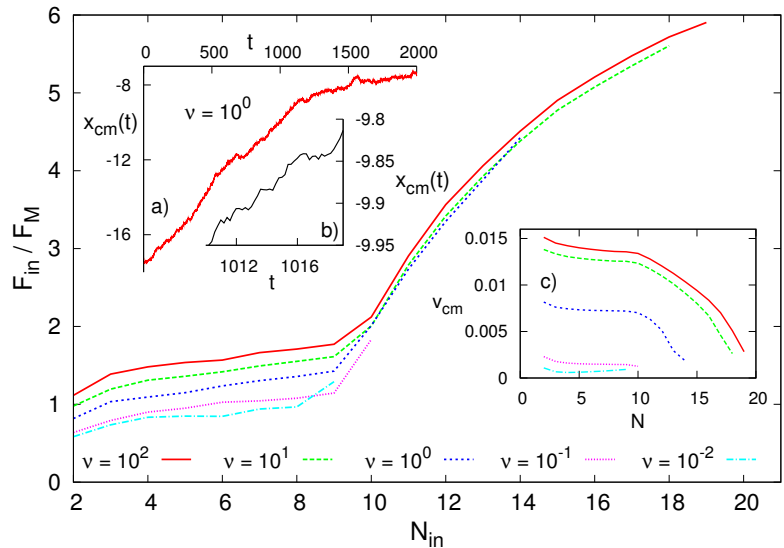}
\caption{Force on the polymer as a function of the
number of monomers entered inside the capsid ($N_{in}$). This curve
resembles the experimental outcome reported in \cite{Bust01}. The
length of the capsid is $L_{Caps}=8$ and the number of monomer
is $N=36$. Insets: a) trajectory of the centre of mass $x_{cm}$ with b) a small portion showing stops; c) chain velocity as a function of $N_{in}$. The activation rate of the motor for the calculations presented in the insets is $\nu=1$.}
\label{Forcecapsid}
\end{figure}

We performed a set of calculations where the chain is pushed into a
limited portion of space. Figure~\ref{caps} shows the scheme of the
motor of length $L_M$ that pushes the polymer inside a region of
length $L_{Caps}$. The polymer starts with the last monomer on the
right at the position $x=0$. This particle ($N$-th monomer in the
figure) cannot pass the wall on the right. Once there, the polymer
begins to compress and the pressure on the particles inside the capsid
increases more and more avoiding in some cases the completion of the
translocation process. To consider excluded volume effects and
maintain the relative order of the monomers, a repulsive only
Lennard-Jones potential has been taken into account in the model:
$V_{LJ}(r)=4\epsilon [ (\frac{\sigma}{r})^{12}- (\frac{\sigma}{r})^6
]$ for $r\leq 2^{1/6}\sigma$, and $0$ otherwise. We used $\epsilon=1$,
and $\sigma=0.1$.

Figure~\ref{Forcecapsid} plots $F_{in}$, the mean force acting on the
last monomer entering in the capsid, as a function of the total number
of monomers entered. The capsid width is $L_{Caps}/d_0=8$ and the
chain length is $N=36$ monomers. It shows that, until the size of the
entered polymer approximately equals the size of the capsid, the force
grows up very slowly. Once the last monomer ($N$) touches the wall of
the capsid, the force inside grows rapidly with a saturating trend at
high number of monomers.  The inset c) of the figure plots the polymer
mean velocity {\it vs} the number of monomers in the capsid. We find a
behaviour similar to that experimentally observed in \cite{Bust01}. It
shows that the model here introduced can depict qualitatively the
packing features of the $\phi 29$ motor despite of its simplicity.

\textit{Biological values.} The model we present here is a simplification of
both a real polymeric chain and a molecular motor:
in the example here used the DNA and the motor of the $\phi 29$ bacteriophage.

The
model is in principle adaptable to any translocation process mediated
by ATP-based motors, provided that a proper scaling of the measurable
quantities can be done. Experimental possible value of the working
time of the motor is $T_M=10 ms$, of the affinity constant $k_M
\backsimeq 30 \mu M$ \cite{Bust09}, and of the saturation velocity
$v_{max} \backsimeq 103$ DNA base pairs/s $= 70 nm/s$
\cite{Bust05}. The saturation velocity sets the $v_{HR}$ value in our
model, and the frequency at the Michaelis-Menten concentration
corresponds to $\nu=\nu_0=1/T_M$. Finally, the stall force value found
experimentally in the high frequency limit is $F_{Stall}^{exp} = 57
pN$ \cite{Bust01}. Using this value we can estimate $F_M \approx 12
pN$. In the same way the energy consumed per cycle can be
  obtained for large ATP concentration as $W = n_{mot}^{on} v_{max}
  F_M T_M$. With the parameters used, $W \approx 12 kT_{room}$, which
  is less than the energy provided by an ATP molecule, i.e. around $20
  kT_{room}$.

\section{Discussion}

In this work we have introduced a simple model for a
molecular motor which consumes ATP and produces mechanical work
pushing a polymer chain.

Other simple models have been used to describe translocation features
of DNA. For instance, in the works by Downton a 1d polymer chain
jointed together by a FENE potential is studied~\cite{linke,linke2}. The
chain moves aided by a flashing extended ratchet potential. By
contrast, our potential acts in a limited region of space which is a
more realistic approach to translocation through a pore.

In our work, we calculate the mean translocation time and the velocity
of the polymer as function of the activation frequency of the
motor. The latter shows a very good agreement with the experiments and
follows a clear Michaelis-Menten dependence with the ATP
concentration.  We see that such MM laws arise in a natural way in the
description of the system as a consequence of the kinetics of a
machine which remains active for a certain given time. It appears that
the {\it working time average} of the molecular motors which use ATP,
is actually the origin of the Michaelis-Menten law in ATP-motor
assisted dynamics.

We have also studied the behaviour of the polymer in the presence of a
pulling force and obtained analytical expressions for the stall force
of the system as a function of the ATP concentration.  Such
expressions, a corrected MM equation, show an excellent agreement to
our computed results.  Finally, the force inside the capsid and the
velocity of the chain as a function of the amount of polymer packed
have been also evaluated finding good qualitative agreement with the
experiments.

We have studied a one dimensional model. Polymer translocation
experiments are usually performed with the help of optical traps. That
way, the polymer (DNA or RNA) is held almost completely stretched
out. With respect to the capsid effect, the confinement introduced in
the model describes a saturating pressure inside the capsid without
taking into account specific geometrical details and polymer recoil
effects. For these two reasons the one-dimensional model is a good
enough first approach to study the process.

Besides the fact that technological improvements allow nowadays very
precise investigations at the nanoscale, the microscopic working
details of the bacteriophage motors are not completely understood.
The model is close to one of the two mechanisms recently proposed in
\cite{Bust09}, where the force acts by means of steric interactions,
without chemical bonds with the DNA. Since DNA packing motor $\phi29$
is a well studied motor protein with several intriguing and
unexplained features we have compared our results to experiments with
this motor. In general we have found a good agreement. However, a
detail model for $\phi29$ should include many other aspects and it is
beyond the spirit and purpose of our work. Between these aspects we
could mention a precise modelling of the inactive stage, the inclusion
of more complex waiting time statistics or the addition of existing
structural data.

The results here presented describe general features of a motor which
actuates with a mean force during its cycle. In this sense the
qualitative results here reported need some adjustments of the
parameters if applied in concrete cases and can be applied to a wide
class of ATP-based motor, independently on the inner functioning of
the very motor. Our work shows that our model could be a good starting
point to develop detailed descriptions of different motors.

\section*{Acknowledgments}
This work was supported by Spanish MICINN through
Project No.FIS2008-01240, cofinanced by FEDER funds.

\end{document}